# Suppressing DC Drift in Thin-Film Lithium Niobate Modulators via Multiferroic Skyrmion Excitation

**Authors:** Yalong Yu[1], Yekai Ren[1], Nuo Chen[1], and Tao Chu[1, *]

**Affiliations:**
[1] College of Information Science and Electronic Engineering, Zhejiang University, Hangzhou 310058, P. R. China

[*]Corresponding author. Email: chutao@zju.edu.cn;

**Abstract:** Thin-film lithium niobate (TFLN) modulators, despite their superior electro-optic performance, face critical DC drift challenges under low-frequency or prolonged operation. In this work, we demonstrate a novel suppression strategy by exciting multiferroic skyrmions in TFLN, achieving drift-free square-wave modulation for >1 hour—the first solution eliminating feedback systems. This breakthrough originates from dual carrier suppression mechanisms: (1) charge density reduction via skyrmion-induced polarization nano-regions (PNRs) excitation, and (2) mean free path restriction through polarization gradients at PNRs domain walls. By directly targeting the root cause of DC drift—mobile charge redistribution—our method uniquely preserves the essential $SiO_2$ upper cladding, resolving the longstanding trade-off between drift mitigation and waveguide protection. Crucially, our work also provides the first experimental observation of interconversion between short-term drift (seconds-scale transient overshoot) and long-term drift (hours-scale baseline shift), offering critical insights into their unified origin.

Recently, TFLN optical modulators have emerged as pivotal components in photonic integrated circuits (PICs) due to their exceptional modulation bandwidth, low optical loss, and high modulation efficiency. However, their practical deployment faces fundamental limitations due to cumulative DC drift effects that progressively deteriorate modulation stability, particularly under low-frequency operation and prolonged DC biasing. Prior research has extensively characterized DC drift mechanisms in TFLN systems, primarily ascribing the phenomenon to charge relaxation dynamics stemming from crystalline imperfections—including intrinsic atomic vacancy migration and interfacial lattice discontinuities at LN-SiO$_2$ heterojunctions (1). While strategic interface modifications have demonstrated partial success in suppressing surface charge accumulation, all existing drift-mitigation protocols paradoxically require complete removal of the protective SiO$_2$ cladding—a sacrificial approach that exposes waveguides to environmental degradation and renders devices incompatible with industrial packaging standards (2~4).

Here, we present a breakthrough approach where we effectively suppress DC drift in electro-optic (EO) modulators through the controlled excitation of high-density multiferroic skyrmions in LN (4). Remarkably, the optimized modulator achieves stable square-wave modulation without observable DC drift under continuous DC voltage operation for over 1 hour. Crucially, our methodology preserves the essential SiO$_2$ cladding layer - the first demonstrated drift mitigation strategy that neither requires cladding removal nor auxiliary feedback systems, thereby circumventing waveguide contamination risks inherent in previous approaches. Notably, we observe interconvertibility between short-term and long-term drift components, suggesting their common physical origin. This magnetic manipulation paradigm establishes a fundamentally new direction for addressing stability challenges in TFLN devices.

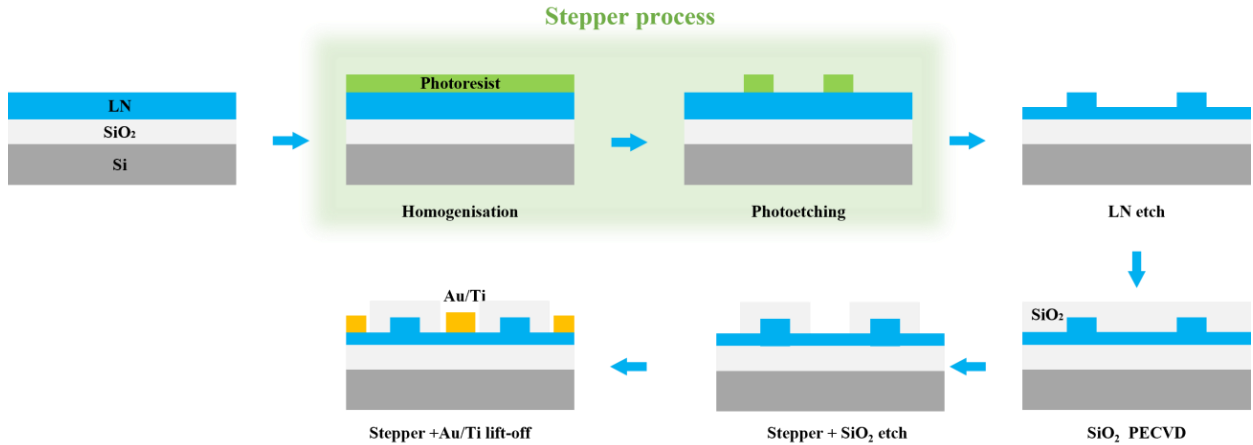

**Fig. 1.** Fabrication processes of the TFLN modulator

In this experiment, fabrication of the modulator was performed on a commercial X-cut TFLN platform consisting of a 600-nm single-crystalline LN layer bonded via 4.7-μm thermal SiO$_2$ to a 525-μm silicon substrate. Waveguide patterning was achieved through 365-nm i-line stepper lithography followed by inductively coupled plasma reactive ion etching (ICP-RIE) with 200-nm etch depth. An 800-nm SiO$_2$ cladding layer was subsequently deposited by plasma-enhanced chemical vapor deposition (PECVD), with selective ICP etching creating recessed regions for electrode integration. Multilayer electrodes comprising 40-nm Ti adhesion layer and 600-nm Au were fabricated via electron-beam evaporation and lift-off processing (Fig. 1).

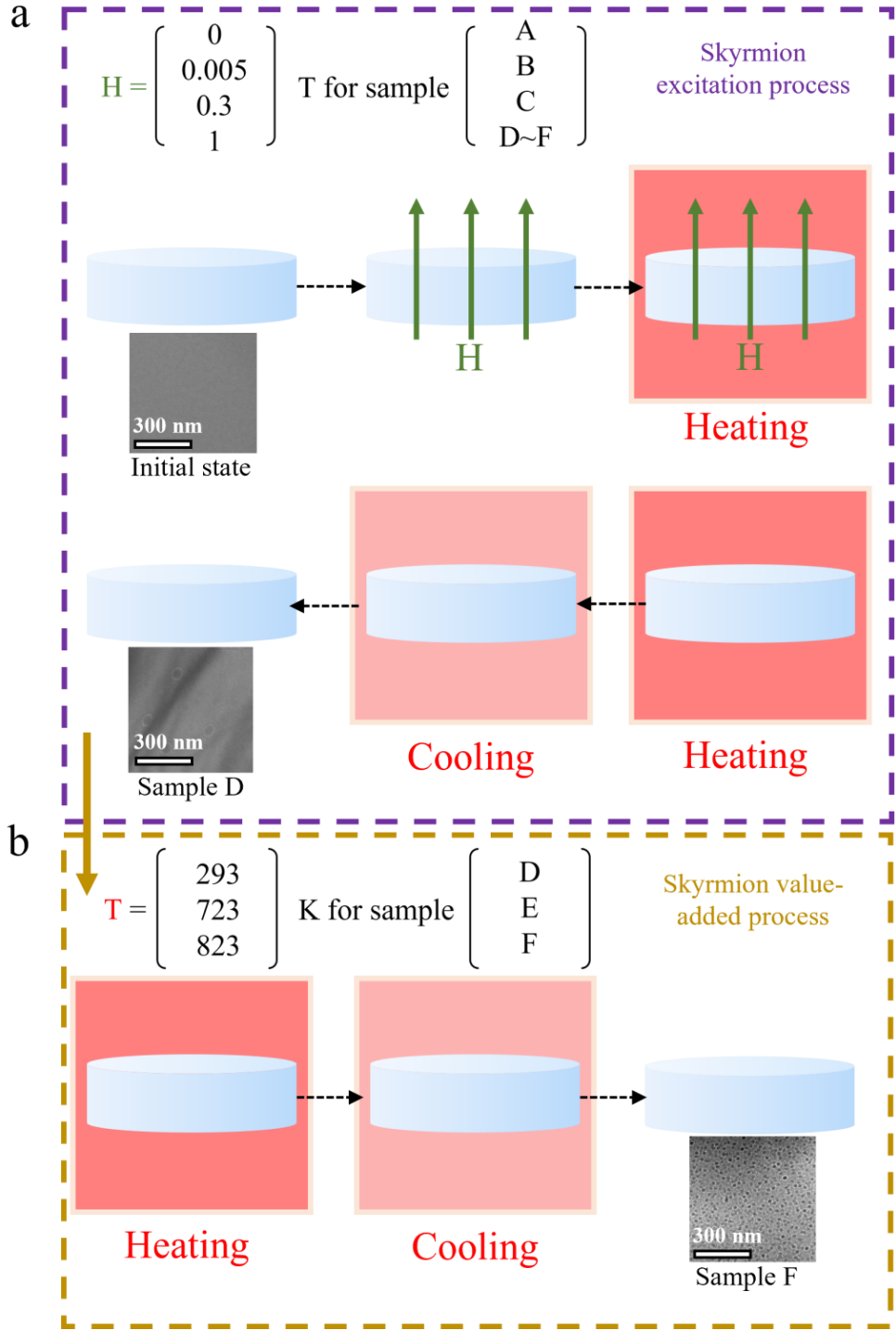

**Fig. 2. Skyrmion excitation and density control in TFLN modulators.** (**a**) Magnetic annealing processes for Samples A, B, C, and D~F under varying external fields (0 T, 0.005 T, 0.3 T, 1 T) at 823 K, with LTEM images showing pristine LN (no skyrmions) versus processed Sample D (sparse skyrmion states). (**b**) Zero-field thermal annealing of Samples D-F at 293 K (no annealing), 723 K, and 823 K, with LTEM imaging of Sample F revealing dense skyrmion state post-823K treatment.

Following device fabrication, a systematic investigation was conducted across five modulator configurations to elucidate the impact of magnetic field processing on DC drift suppression: Sample A (untreated control) provided baseline performance, while Samples B-F underwent progressively intensified magnetic annealing protocols. Samples B-F were subjected to a four-phase thermal-magnetic sequence comprising (1) initial magnetic field application at ambient conditions (H = 0.005, 0.3, and 1 T respectively), (2) controlled temperature ramp (100 K/min) to 743 K with sustained magnetic exposure for 30 min, (3) in-situ magnetic field termination followed by 30 min thermal equilibration, and (4) passive cooling to room temperature. Sample E, F uniquely incorporated a supplementary zero-field annealing step at 773K and 823 K for 30 min following Group D processing, specifically designed to amplify skyrmion nucleation density through thermal reorganization. Lorentz transmission electron microscopy (LTEM) analysis confirmed these two critical transitions: (i) magnetic-field-induced skyrmion excitation in processed devices (Fig. 2a), with density increasing from $0/\mu m^2$ (pristine LN) to $50/\mu m^2$ (Sample D, 1 T processing); and (ii) fieldless-annealing-driven density amplification to over $4000/\mu m^2$ (Sample F, 823 K treatment) as shown in Fig. 2b.

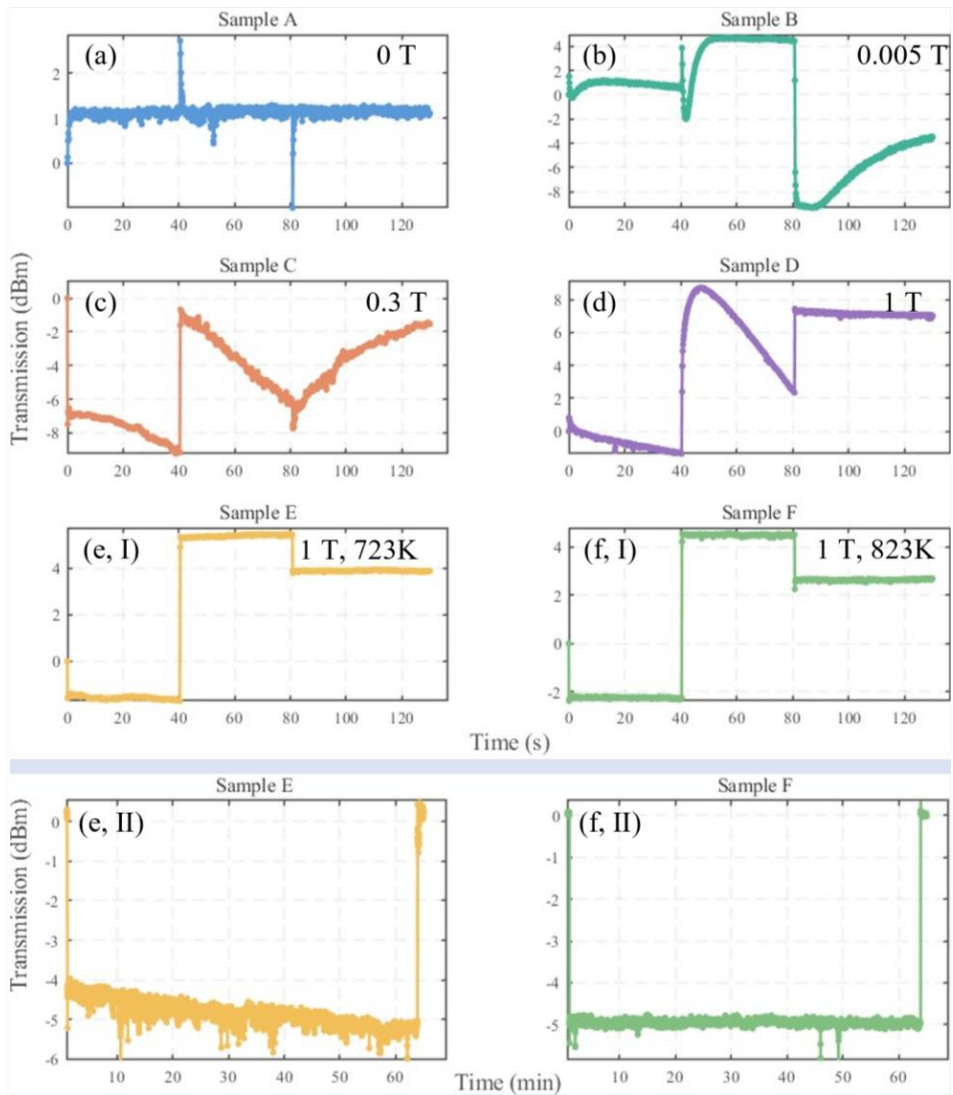

**Fig. 3. DC drift test results of the samples under applied DC voltage**. During the short-term drift test, 4V instantaneous bias is applied at 0~40s, -4V instantaneous bias is applied at 40s~80s, and the voltage is adjusted to 0 at 80~120s. In the long-term drift test, an 8V bias voltage is applied at 0s and maintained for more than an hour, after which the bias voltage is instantly reduced to 0.

We employed a precisely timed square-wave voltage sequence synchronized with optical power monitoring to evaluate the dc drift of the modulator: Initial 0 V at 0s established baseline equilibrium, followed by +4 V application (0-40 s), subsequent polarity reversal to -4 V (40-80 s), and final return to 0 V (80-120 s). This cyclic excitation profile enabled simultaneous characterization of transient response characteristics during voltage transitions and long-term stabilization phenomenon under sustained bias.

Control Sample A (Fig. 3(a)) exhibited severe short-term drift with an instantaneous signal overshoot upon +4 V/-4V application, decaying to near-zero residual drift within 1 s. Magnetic processing fundamentally altered these dynamics: Sample B (Fig. 3(b)) demonstrated reduced overshoot with emergent bimodal relaxation (fast $\tau_1 = 1$ s, slow $\tau_2 = 10$ s components), indicating partial conversion of short-term drift into long-term drift. With progressive intensification of the magnetic field (Fig. 3(c, d)), the transient drift components exhibited monotonic suppression while long-term drift became increasingly pronounced. This parametric evolution induced a fundamental transition in drift dynamics—the dominant drift mechanism transitioned from rapid transient overshoot ($\tau_1 < 1$ s) to gradual baseline migration ($\tau_2 > 10$ s).

Through additional annealing to enhance skyrmion density, we observed continued suppression of DC drift. After 30-minute annealing at 723 K (Fig. 3(e)), the previously intensified long-term drift exhibited significant attenuation, with optical signals transitioning into well-defined square waveforms—direct evidence of skyrmion-mediated drift suppression. Under 8 V DC bias, the 1-hour long-term drift decreased to ~1 dB, comparable to results from $SiO_2$-cladding-removed counterparts. Further increasing the annealing temperature to 823 K yielded drastically weakened dc drift (Fig. 3(f)): Transient overshoot during bias application reduced to <0.3 dB, while 1-hour long-term drift diminished to <0.1 dB—on par with the system's inherent signal noise level. This indicates near-complete DC drift elimination in Sample F.

Our magnetic processing and supplemental annealing protocols achieved drift suppression performance rivaling both cladding-removed TFLN devices and thin-film lithium tantalate modulators. Throughout this process, we documented the conversion of short-term drift into prolonged temporal evolution—a phenomenon contradicting conventional theories that ascribe independent origins to these drift components. This finding exposes critical limitations in existing models of lithium niobate's DC drift mechanisms.

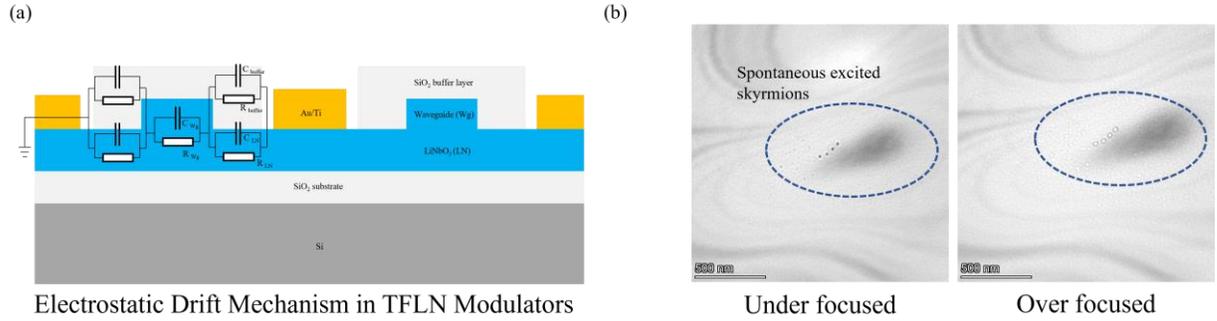

**Fig. 4. Mechanism of skyrmions-induced dc drift reduction**. (**a**). Acknowledged electrostatic model of dc drift in TFLN. (**b**) LTEM images of spontaneously formed skyrmion.

After achieving significant suppression of DC drift in TFLN modulators through multiferroic skyrmion excitation, we use our experimental results to reinterpret the microscopic mechanisms behind DC drift. First, we consider the widely accepted equivalent capacitance model (Fig. 4(a)). In this model, the DC drift is entirely attributed to contributions from unbound charges at the interfaces of LN, $SiO_2$, and electrodes. Upon bias application, the heterogeneous material stack (LN thin film, $SiO_2$ cladding, and electrode interfaces) forms a distributed RC network governed by the time constant $\tau = RC$. This fundamental relationship dictates the temporal evolution of charge redistribution—reducing either resistance (R) or capacitance (C), for example, can significantly accelerate the transient response ($\tau \downarrow$), thereby intensifying DC drift amplitude within operational timescales.

Here, we propose that the observed weakening of long-term drift may arise from two distinct mechanisms:

1. Reduced response time: The refractive index changes caused by drift are accelerated, shifting from timescales of days/months to minutes/seconds.

2. Prolonged response time: The drift response is stretched temporally, meaning a drift level previously reached in time $t$ now requires $N \times t$, flattening the drift signal and making it negligible within practical operational periods when $N$ is sufficiently large.

Previous studies have documented analogous charge acceleration phenomena (Mechanism 1), where plasma bombardment-induced lattice disordering generates excess free charges, effectively reducing the RC time constant ($\tau = RC \downarrow$) and converting long-term drift into accelerated transient responses ([3](#)). In this work, however, our experimental data provide conclusive support for the second mechanism, as evidenced by the progressive flattening of DC drift profiles with increasing treatment intensity (Fig. 3). This behavior signifies a substantial increase in either the effective resistance (R ↑) or interfacial capacitance (C ↑). However, it is evident that we did not remove the $SiO_2$ cladding, and the excited skyrmions are three-dimensional structures existing within the LN material. This confirms that the former (increase in effective resistance) is the source of the weakened drift.

It can be intuitively understood by two mechanisms induced by multiferroic skyrmions excitation: When skyrmions are excited in LN, polarization domains form within the material, effectively trapping a large number of free charges, equivalent to a substantial increase in $R$ within the RC circuit model. Direct evidence for this is our observation of spontaneously formed skyrmion states in inhomogeneous regions of the LN film, without magnetic field excitation, indicating that high densities of free charges create favorable conditions for skyrmion formation (Fig. 4(b)). Additionally, dense, localized PNRs restrict directional movement of free charges, significantly slowing charge accumulation at interfaces and reducing the oscillation amplitude per unit time. Consequently, the significant reduction in free charge density and the mean free path restriction through polarization gradients at PNRs domain walls naturally explains the suppression of DC drift.

However, this purely electrical interpretation remains incomplete. A critical unresolved issue is that a 0.005 T magnetic field—insufficient to directly excite skyrmions in LN—still induces measurable drift suppression. We propose that LN's magnetoelectric coupling properties, such as electric-magnetic-electric oscillations or interfacial Dzyaloshinskii-Moriya interactions, may also contribute to this phenomenon. While the complete theoretical framework for DC drift requires further development, our magnetic processing strategy provides a groundbreaking solution for TFLN-based PICs. Importantly, all evidence suggests that incorporating LN's magnetic properties is the most effective pathway for resolving DC drift.

Furthermore, our device processing employed a maximum temperature of 823 K—the upper limit to prevent $SiO_2$-LN interface degradation and excessive optical loss. Higher annealing temperatures could theoretically enable even more stable drift elimination, though constrained by material stability. Crucially, integrating this magnetic treatment during LN thin-film growth (rather than post-fabrication) offers the most viable path for industrial adoption. Our high-temperature magnetization protocol is inherently compatible with mass production, as magnetic field application—unlike electric field patterning—requires no dedicated nano-structuring or stringent field uniformity, significantly simplifying manufacturing workflows.

This paradigm shift toward material-centric solutions—optimizing LN's magnetic properties during crystal growth—represents the most rational direction for achieving zero DC drift in TFLN platforms. By engineering oxygen vacancies to amplify DMI effects, future work could achieve near-complete drift suppression while retaining full compatibility with industrial PIC fabrication standards.

## References

1. K. Steven, An RC Network Analysis of Long Term Ti:$LiNbO_3$ Bias Stability, *J. Lightwave Technol.* **14**, 2687-2697 (1996).

2. Y. T. Xu, M. H. Shen, J. J. Lu, J. B. Surya, A. S. Ayed, H. X. Tang, Mitigating photorefractive effect in thin-film lithium niobate microring resonators, *Opt. Express.* **29**, 5497-5504 (2021).

3. J. K. Shi, Z. L. Ye, Z. Liu, Z. Yan, K. P. Jia, L. Q. Zhang, D. H. Ge, S. N. Zhu, Alleviation of DC drift in a thin-film lithium niobate, *Opt. Lett.*, **50**, 1703-1706 (2024).

4. J. Holzgrafe, E. Puma, R. Cheng, H. Warner, A. Shams-ansari, R. Shankar, M. Loncar, Relaxation of the electro-optic response in thin-film lithium niobate modulators, *Opt. Express.* **32**, 3619-3631 (2024).

5. Y. L. Yu, et al, Stable room-temperature multiferroic skyrmions in lithium niobate with enhanced Pockels effect, https://arxiv.org/pdf/2407.05349